\documentclass[%
 reprint,
amsmath,amssymb,aps,
]{revtex4-1}

\usepackage{graphicx}
\usepackage{dcolumn}
\usepackage{bm}

\usepackage{bm}

\begin{document}

\title{Beam Size Reconstruction from Ionization Profile Monitors}

\author{Vladimir Shiltsev}
\affiliation{Fermilab, PO Box 500, MS339, Batavia, IL 60510,USA}
\date{\today{}}

\begin{abstract}
Ionization profile monitors (IPMs) are widely used in particle accelerators for fast diagnostics of high energy beams. Due to the space-charge effects and several other physical reasons, as well as due to instrumental effects, the measured IPM profiles can significantly differ from those of the beams. There are several empirical mathematical models commonly used for reconstruction of the beam profiles. Here we present a proper correction algorithm based on the space-charge dynamics of the secondaries in IPMs. We also demonstrate the efficiency of the proposed beam size reconstruction algorithm from the measured IPM profiles and discuss practical aspects limiting the IPM accuracy.
\end{abstract}


\maketitle


\section{Introduction}
\label{intro}
\begin{figure}[htbp]
\centering
\includegraphics[width=0.99\linewidth]{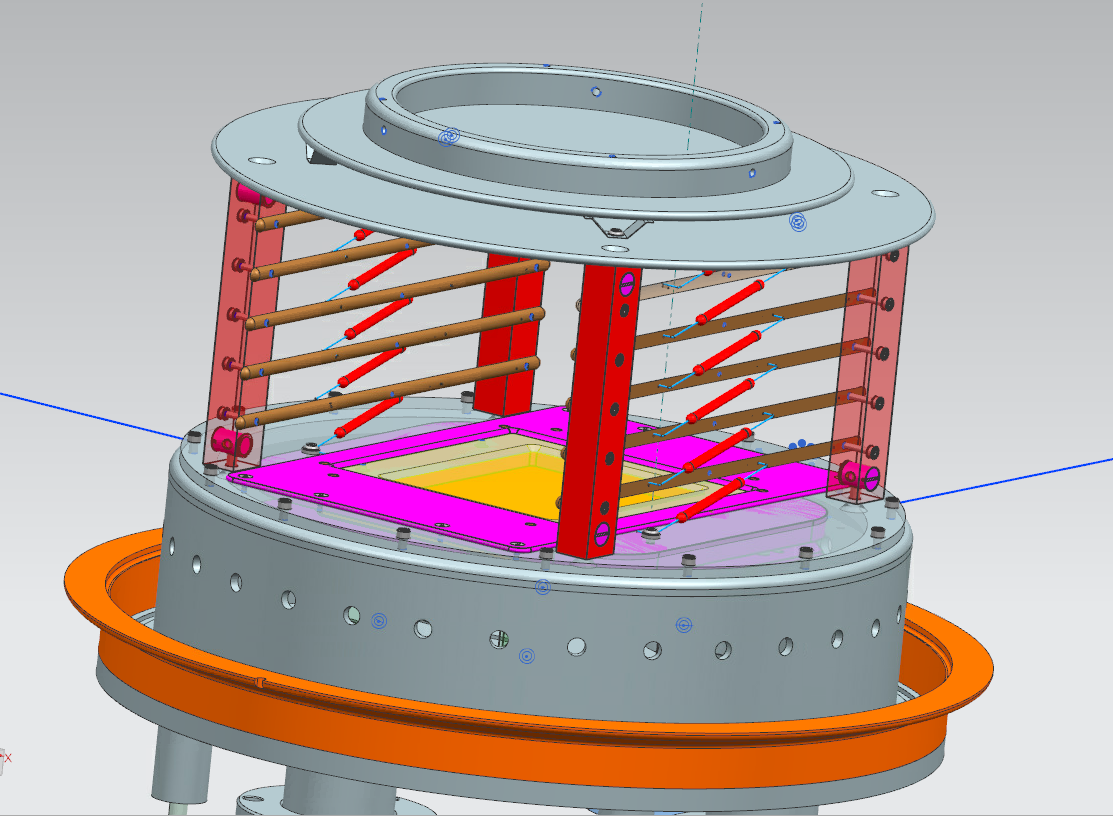}
\caption{Fermilab Booster IPM. Several mm wide proton beam with average current of about $J_p \sim 0.4$A goes from left to right through a 103 mm high HV cage. The maximum voltage on the upper plate is $V_0$=+24 kV, the electric field uniformity is arranged by six-stage voltage divider bars. Secondary ions are accelerated toward a 80$\times$100 mm$^2$ micro-channel plate (MCP, shown in gold). Electrons, exiting from the MCP, proceed for another 7.5 mm to an array of thin, parallel anode strips spaced 1.5 mm apart at +100V above the exit of the MCP (not shown), where the electrons are collected and amplified for further processing (courtesy R.Thurman-Keup).}
\label{Fig.1}
\end{figure}

Ionization profile monitors (IPMs) have been in active use in particle accelerators since the late 1960s \cite{hornstra1967nondestructive, weisberg1983ionization, hochadel1994residual, anne1993noninterceptive, connolly2000beam, jansson2006tevatron, levasseur2017development} and are important beam diagnostic tools for many modern and future accelerators \cite{moore2009beam, benedetti2020design}. Their principle of operation is based on collection of the products of ionization of residual gas by high energy charged particle beams 
They operate by collecting the products created from the ionization of residual gas by high energy charged particle beams - see detailed discussions and examples of operational instruments  in Refs.\cite{strehl2006beam, wittenburg2013specific}. Transverse profiles of the secondaries give a very good approximation of the primary beam properties and usually can be quickly measured on very short time scales, e.g., on a turn-by-turn or even on a bunch-by-bunch basis. The two most common types of IPMs are distinguished by the use or non-use of a guiding magnetic field parallel to the extracting electric field. Physics principles, advantages and disadvantages of the IPMs with a magnetic field are discussed in \cite{vilsmeier2019space}. This paper deals mostly with the physics principles and beam profile reconstruction in the IPMs operating without magnets, with only an electric guiding field. These IPMs are used more widely because of smaller size, simpler design and lower cost (see Fig.\ref{Fig.1}).  

One of the key challenges for the initial beam profile reconstruction is properly accounting for various effects which lead to distortion of the charge distribution of secondaries as they travel  to the detector and in the detector itself. In high-intensity accelerators, the dominant effect is the increase of the measured beam size $\sigma_{m}$ compared to the initial beam size $\sigma_{0}$ caused by space-charge forces of the primary beam. There is an extensive literature on this effect; many simulation codes are developed, presented and discussed in, e.g., proceedings of recent Workshops \cite{workshop2016, workshop2017}. Several empirical mathematical models were proposed to relate $\sigma_{m}$ and $\sigma_{0}$, such as - from \cite{weisberg1983ionization}: 
\begin{equation}
    \sigma_m = \sqrt{\sigma_0^2+ C_1 \frac{N^2}{E_0^2 \sigma_0}} \; ,
\label{eq:fit1}
\end{equation}
or, alternatively, from \cite{thern1987space}: 
\begin{equation}
    \sigma_m = \sigma_0 +C_2 \frac{N^{1.025}}{\sigma_0^{1.65}}(1+1.5R^{1.45})^{-0.28} \; ,
\label{eq:fit2}
\end{equation}

or, from \cite{Graves}: 
\begin{equation}
    \sigma_0 = C_1 + C_2 \sigma_m + C_3 N \; ,
\label{eq:fit3}
\end{equation}

or, from \cite{amundson2003calibration}: 
\begin{equation}
    \sigma_m = \sigma_0+ C_3 \frac{N}{\sigma_0^{0.615}} + C_4 \frac{N^2}{\sigma_0^{3.45}}\; ,
\label{eq:fit4}
\end{equation}
here, $N$ is the number of particles in the high energy primary beam, $E_0=V_0/D$ is the guiding electric field due to the voltage gradient $v_0$ across the IPM gap $D$, $R$ is the aspect ratio of (other plane)/(measured plane), e.g., $R_x=\sigma_{0,y}/\sigma_{0,x}$ for horizontal plane, and $C_1, C_2, C_3, C_4$ are the constants derived to fit available simulations and measurements data. 

Despite acceptable data approximation, the wide variety of mathematical constructs and the unclear physical reasons for various exponents in Eqs.(\ref{eq:fit1} - \ref{eq:fit4}) and, therefore, undefined applicability ranges, are generally confusing and call for either a better analysis or more sophisticated beam profile reconstruction algorithms. 

In the next Chapter we develop a new algorithm based on a well defined physics description and an analysis that results in a complete understanding of the IPM signal dependencies on all major parameters, such as high-energy beam intensity $N$ and size $\sigma_0$, the IPM voltage $V_0$ and dimensions, etc. In the last Chapter we successfully apply our theory to determine the sought-for beam size $\sigma_0$ from the profiles measured by the operational Fermilab Booster proton synchtrotron IPM profiles.  

\section{Space-charge driven IPM profile expansion}

\begin{figure}[htbp]
\centering
\includegraphics[width=0.99\linewidth]{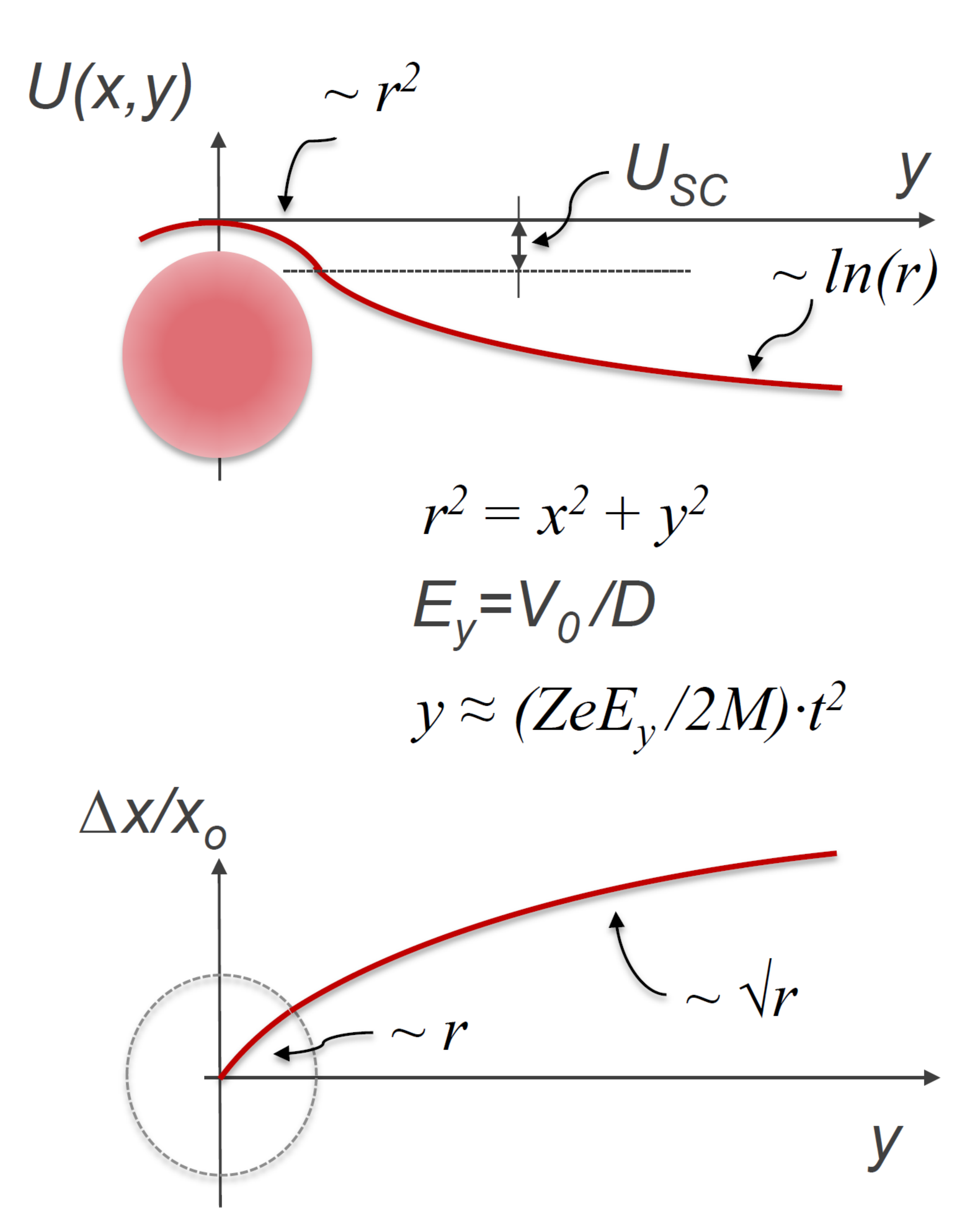}
\caption{Proton beam space-charge effect in IPM: (top) the space-charge potential; (bottom) space-charge driven expansion in $x$-plane in the presence of a much stronger IPM extraction field $E_y$ (see text).} \label{Fig.2}
\end{figure}

The general equations of transverse motion of the charged secondary particles (ions, electrons) born in the IPM in the acts of ionization of the residual gas molecules are 
\begin{eqnarray}
\frac {d^2 x(t)}{dt^2}  & = & f_x(x, y, t) \, x 
\nonumber \\ 
\frac {d^2 y(t)}{dt^2}  & = &  \frac{Ze}{M}E_y + f_y(x, y,t) \, y \; ,
\label{E1}
\end{eqnarray}
where $Ze$ and $M$ are charge and mass of the secondaries, $E_y=V_0/D$ is the IPM extracting external electric field which is assumed to be generated by application of high voltage $V_0$ over the gap $D$, and functions $f_{(x,y)}(x,y,t)= - (Ze / M)\cdot \partial{^2 U(x,y,t)}/ \partial {(x,y)^2}$ reflect the space-charge impact of the primary high energy particle beam. The space-charge potential $U(x,y, t)$ is proportional to beam current $J(t)$ and depends on the beam density distribution. For typical transverse beam current distributions in accelerators it scales as $\propto r^2=x^2+y^2$ at distances less than a characteristic beam size $a$ and as $\propto$ ln$(r)$ for $r \gg a$, as schematically shown in Fig.\ref{Fig.2}. 

For initial analysis we omit complications due to the field distortions at the IPM boundaries (such as grounded potential at the MCP plane), assume DC beam current $J$ and for the simplest case of uniform beam with radius $a$ one gets:
\begin{eqnarray}
    U(x,y) & = & - U_{SC} \frac{r^2}{a^2} \; \, \, \, {\rm for} \,  r<a \; 
    \nonumber \\
    & = & - U_{SC} (1+2\,{\rm ln} (r/a)) \; \, \, \, {\rm for} \, r \ge a \; 
\label{potential}
\end{eqnarray}
where $U_{SC}=J/(4 \pi \epsilon_0 v_p) \approx 30[V/A]J/\beta_p$, $\beta_p=v_p/c$,  $v_p$ the main (proton) beam velocity, $c$ is the speed of light, and $\epsilon_0$ is the permittivity of vacuum \cite{Reiser}. The analysis can be further simplified by taking into account that not only is the space-charge potential $O$(10 V) usually small compared to $O$(10-100 kV) IPM voltage $U_{SC} \ll V_0$, but its gradient $\sim U_{SC}/a$ which is $O$(10 V/mm) in its peak at the edge of the beam is also small compared to the uniform IPM electric field $E_y$ that is $O$(100 V/mm). In this case, the equation of motion in the $y$-plane becomes trivial: 
\begin{equation}
    y(t) = \frac{ZeE_y}{2M} t^2 + v_{0,y}t+y_0 \; ,
\label{equationY}
\end{equation}
where $y_0$ and $v_{0,y}$ are the original position and velocity of the secondary particle at the moment of its creation. Combination of the last three equations makes the equation of motion in the plane of expansion as
\begin{equation}
   \frac {d^2 x(t)}{dt^2}= \frac{x} {\tau_1^2} \; ,
\label{equationX1}
\end{equation}
for particle trajectories inside the high energy beam $r(t)<a$, while outside the high energy beam $r(t)=\sqrt{y(t)^2+x(t)^2} \ge a$ it is: 
\begin{equation}
   \frac {d^2 x(t)}{dt^2}= \frac{x} {\tau_1^2} \frac{a^2} {r(t)^2} \; .
\label{equationX2}
\end{equation}
Here we introduced a characteristic expansion time due to the space-charge: 
\begin{equation}
   \tau_1 = \Big( \frac{2eZU_{SC}} {Ma^2} \Big)^{-1/2} \; . 
\label{tau1}
\end{equation}

The structure of the second-order ordinary differential equations (\ref{equationX1}, \ref{equationX2}) is such that the particle's transverse velocity is mostly accumulated while it is passing through the beam core area, while the beam  trajectory outside it is mostly ballistic and scales approximately linearly with time until the particle reaches the IPM detector (MCP) at a $t=\tau_2$: 
\begin{equation}
   \tau_2 = \sqrt{\frac{2MDd} {ZeV_0} } \; , 
\label{tau2}
\end{equation}
here $d$ is the average distance from the beam center to the MCP plate. 
Solution of equation (\ref{equationX1}) is a  {\it hyperbolic} function \cite{Morse}:
\begin{equation}
    x(t) = x_0 {\rm ch}(t/\tau_1) + v_{0,x}\tau_1 {\rm sh}(t/\tau_1) \; .
\label{equationX3}
\end{equation} 
It holds while the particle is inside the proton beam, $r(t) \leq a$, that corresponds to $t\leq \tau_0$ where
\begin{equation}
   \tau_0 = \sqrt{\frac{2eMDa} {ZeV_0}}=\tau_2 \sqrt{a \over d} \; , 
\label{tau0}
\end{equation}
is a characteristic time for hte secondaries to get extracted out of the beam by the external electric field $E_y=V_0/D$.  For an initial approximation one can assume that the secondary particles are born with negligible initial kinetic energy $\mathcal{E}_i=Mv_0^2/2 \ll (eV_0, eU_{SC})$ (see more in the next section), and the second term in Eq.(\ref{equationX3}) can be neglected. Together with the smallness of the initial coordinates compared to the average distance $d$ from the beam center to the MCP plate $(x_0, y_0) \ll d \approx D/2$, one gets for $r(t)> a$ using approximation $r(t) \approx y(t) =  t^2 ZeE_y/2M$ in the denominator of Eq.(\ref{equationX2}) : 
\begin{equation}
    x(t) = x_0 \frac{t}{\tau_0} \Big[ {\rm ch}(\alpha)\big({\rm ch}(\tilde{\alpha})-{{\rm sh}(\tilde{\alpha}) \over \alpha}\big) + {\rm sh}(\alpha) {\rm sh}(\tilde{\alpha}) \Big] \, ,
\label{equationX4}
\end{equation} 
where $\alpha=\tau_0/\tau_1$ and $\tilde{\alpha}=\alpha(1-\tau_0/t)$. The above solution matches Eq.(\ref{equationX3}) at the boundary, i.e., $x(\tau_0)=x_0{\rm ch}(\tau_0/\tau_1)$ and $x'(\tau_0)=x_0/\tau_1 {\rm sh}(\tau_0/\tau_1)$. Fig.\ref{Fig.2} shows the main qualitative features of the space-charge driven particle trajectory's expansion. For $t \gg \tau_0$ and $\tau_0 < \tau_1$, the leading terms of the solution are:
\begin{equation}
    x(t) \approx x_0 \Big[ 1 + {4 \over 3} \frac{t \tau_0}{\tau_1^2} 
 \big( 1+{2 \over 5} \frac{\tau_0^2}{\tau_1^2} + ... \big) \Big] \, .
\label{equationX5PR}
\end{equation} 

Given that Eqs.(\ref{equationX1}) and (\ref{equationX2}) are linear in $x$, it is no surprise that the transformation Eq.(\ref{equationX5PR}) is linear with respect to $x_0$, too, leading to {\it proportional magnification of the profile} of the distribution of the secondary particles. Accordingly, the rms transverse size of the IPM profile at the time $\tau_2$ when the secondary particle reaches the MCP detector becomes:   
\begin{equation}
    \sigma_m = \sigma_0 \cdot h \approx \sigma_0 \cdot \Big(
1+ {4\sqrt{2} \over 3} \frac{U_{SC}D}{V_0 \sigma_0} \sqrt{d \over \sigma_0} 
\Big) \, ,
\label{equationX6}
\end{equation} 
where $\sigma_0=a/2$ is the primary beam horizontal rms size and we retain only the first two terms in Eq.(\ref{equationX5PR}).  It is remarkable that the space-charge expansion factor $h$ is determined only by the space-charge potential $U_{SC}$, the primary beam size $\sigma_0$, the IPM extracting field $E_y=V_0/D$, and the beam-to-MCP distance $d$ but {\it it does not depend on the type of secondary species} (their mass and charge, etc). The basic reason is that both the space-charge impact along $x$-axis and the transport mechanism along $y$-axis are set by electric fields. That condition would not hold if, for example, the particle had significant initial velocity $v_{0,y}$ and the second term dominated in the right-hand side of Eq.(\ref{equationY}). 

A similar analysis for Gaussian, rather than uniform, proton beam current density distribution with rms size $\sigma_0$ yields essentially the same result, with the numerical factor $(4\sqrt{2}/3)\approx 1.88$ replaced by $2\Gamma(1/4)/3\approx 2.41$, and substitution of $\sigma_0=a/2$ - see Appendix \ref{A1}. In general, Eqs.(\ref{equationX5PR}) and (\ref{EQ5}) correctly describe the dynamics of the majority of the secondary particles, originating from the core of the proton beam $(x_0, y_0) \leq \sigma_0$. The remaining ones are subject of nonlinear forces but only for the initial part of their trajectories, while their dynamics for the remaining part when $y(t)>(x_0, y_0)$ is still described by linear Eq.(\ref{equationX2}). Accounting for that effect, as well as for the additional acceleration in $y$-plane due to the space-charge force (and, therefore, slightly shorter times $\tau_0$ and $\tau_2$) leads to small next-order corrections proportional to the square of the small parameter $(\tau_0/\tau_2)^2$ and can be neglected for the most practical cases. The effect of the high energy beam size aspect ratio $R=\sigma_y/\sigma_x$ is relatively weak. Indeed, the space-charge factor $1/\tau_1^2$ in Eqs.(\ref{equationX1}) and (\ref{equationX2}) scales as $2/(1+R)$ while the characteristic time $\tau_0 \propto \sqrt{R}$, therefore the product $t\tau_0/\tau_1^2$ $-$ see Eq.(\ref{equationX6}) $-$ will be scaled by $2 \sqrt{R}/(1+R)$. The corresponding correction is 0.94 for $R=0.5$, i.e., relatively small and can be safely neglected for most common cases of $h \leq 2$. As we will show in the next Chapter, Eq.(\ref{equationX6}) can be easily solved, and the original $\sigma_0$ can be found from $\sigma_m$ with known IPM parameters and the beam orbit position $d$.   
The space-charge expansion with a typical factor $h(\tau_2) \leq 2$ is the largest, though not the only one, of the systematic IPM errors. The most important effects to be taken into account in that regard are: a) the time structure of the high-energy beam current $J(t)$, especially with high bunching factor $B=J_{peak}/J_{avg}$; and b) the effect of initial (thermal) particle velocities $v_{0 (x,y)}$.   

The effect of the beam current time structure, such as in bunched beams, depends on the rms bunch length $\tau_b$ and time between bunches $t_b$. In the case $\tau_b \ll t_b  \lesssim \tau_0 \ll (\tau_1,  \tau_2)$, the dynamics of the cloud of secondary particles in IPM is set by a sequence of frequent kicks instead of smooth functions as in Eqs. (\ref{E1}, \ref{equationX1}, \ref{equationX2}) but that does not change the integrals Eqs.(\ref{equationX5PR}, \ref{EQ5}).  Of course, these equations are also applicable in the case of very long bunches $\tau_b \gg \tau_2$. In the case when bunch spacing $t_b$ gets comparable to $\tau_0$ and $\tau_2$, our algorithm should employ summation rather than integration and that leads to the systematic increase of the numerical coefficients in Eqs.(\ref{equationX6}, \ref{EQ6}) by a factor of $U_{SC} \rightarrow U_{SC}(1+0.8\, t_b/\tau_0)$ (see Appendix \ref{A2}).    

Most significant deviations from the above analysis will take place in the case of short and rare bunches $\tau_b \ll (\tau_0, \tau_1,  \tau_2) \ll t_b$. In that case, the position of the secondary particle (ion) remains unchanged during the passage of the bunch, and its dynamics  is all set by almost instantaneous impact (velocity change) following the act of ionization: 
\begin{equation}
\Delta v_x = \frac{2Ze^2N_p}{(4 \pi \epsilon_0) \beta_p Mc} \frac{x_0}{r_0^2} \Big( 1- {\rm exp} (-{r_0^2 \over {2 \sigma_0^2}}) \Big) \, , 
\label{short1}
\end{equation} 
where $N_p$ is the total number of particles in the Gaussian proton bunch, which passed by the ion located at $(x_0,y_0)$, and $r_0^2=x_0^2+y_0^2$. After this impact, the secondary (ion) sees no transverse field $E_x=0$ and proceeds to the IPM collector plate under the extracting field $E_y$. After corresponding time $\tau_2$, thus the resulting displacement $x_0+\tau_2 \Delta v_{x}$ becomes equal to: 
\begin{equation}
x(\tau_2) = x_0 \Big[ 1+ 4 \kappa {N_p \over r_0^2} \Big( 1- {\rm exp} (-{r_0^2 \over {2 \sigma_0^2}}) \Big) \Big] \, ,  
\label{short2}
\end{equation} 
where 
\begin{equation}
\kappa=\sqrt{\frac{eZdD}{(4 \pi \epsilon_0) 2M\beta_p c V_0}} \, .  
\label{short2a}
\end{equation} 

Averaging over a 2D Gaussian distribution of initial positions and taking into account that on average freshly generated ions experience the impact of only half of the bunch, one gets the rms size of the IPM profile: 
\begin{equation}
\sigma_m^2 = \sigma_0^2 + \kappa N_p + \kappa^2 {N_p^2 \over {\sigma_0^2}} {4 \over 3} \ln(4/3) \, ,  
\label{short3}
\end{equation} 
- see Appendix \ref{A3}.
Contrary to the case presented in the preceding section and summarized in Eq.(\ref{equationX6}), the intensity dependent profile expansion of Eq.(\ref{short3}) manifests as an addition in quadrature and it now depends on charge $Z$ and mass $M$ of the secondary species. In particular, to minimize such expansion in the IPM measuring profiles of short intense and rare bunches, it is beneficial to collect (heavier) ions instead of (light) electrons.

Another effect that calls for the use of ions rather than electrons in IPMs without external magnetic fields is caused by the initial velocities of the secondaries $v_0$. Indeed, assuming such velocities are random with the rms value of $\sqrt{2\mathcal{E}_i/M}$, one gets in quadrature addition to Eq.(\ref{equationX6}):    
\begin{equation}
\sigma_m^2=\sigma_0^2 h^2(U_{SC},\sigma_0, V_0, D,d) + 
\Big( \frac{4\mathcal{E}_idD}{ZeV_0}\Big) \, .
\label{Sigma5}
\end{equation} 
At face value, this additional term is independent of the mass of the secondary particle, but the initial kinetic energy strongly depends on the collected species. For example, for ionization electrons, $\mathcal{E}_i$ is about 35 eV needed on average for ion-electron pair production by protons in hydrogen \cite{bakker1951stopping}, and, therefore, the corresponding smearing of the particle position measured by the IPM is about $\sigma_T=D \sqrt{2\mathcal{E}_i/ZeV_0}$, that is some 6 mm for voltages as high as $V_0=20$ kV and a typical $D=100$ mm. That is absolutely unacceptable for millimeter-scale or smaller beam sizes. Electron-collecting IPMs do have and advantage of very short electron reaction time $\tau_2$ and , therefore, excellent time resolution, but they must have the external magnetic field $B_y$ to suppress the smearing. As for ions, their initial kinetic energy depends on their kind and the type of reaction. For diatomic gases, the most relevant process is dissociative ionization by the primary fast protons, i.e., $p+H_2 \rightarrow p+H+H^+$ with typical kinetic energy of the $H^+$ of the order of a few eV \cite{Dimopoulou2005}. Corresponding smearing $\sigma_T$ can reach 1 mm or more \cite{Egberts2012}.


\section{Application for Fermilab Booster}

Below we will apply our analysis to the IPM profiles measured in the Fermilab Booster rapid cycling synchrotron (RCS) \cite{Shiltsev2020}. The Fermilab Booster \cite{Booster, BoosterBook} is a 474.2 m circumference, alternating-gradient 15 Hz RCS accelerating protons from 0.4 GeV at injection to 8.0 GeV at extraction in 33.3 ms, or about 20 000 turns $-$ half of the magnet cycle period. Correspondingly, all proton beam parameters (intensity, positions, bunch length, emittances) as well as accelerating RF frequencies and voltage significantly vary over the cycle. Typical total intensity of 84 circulating proton bunches is about $N=4.6 \cdot 10^{12}$. 

\begin{figure}[htbp]
\centering
\includegraphics[width=0.99\linewidth]{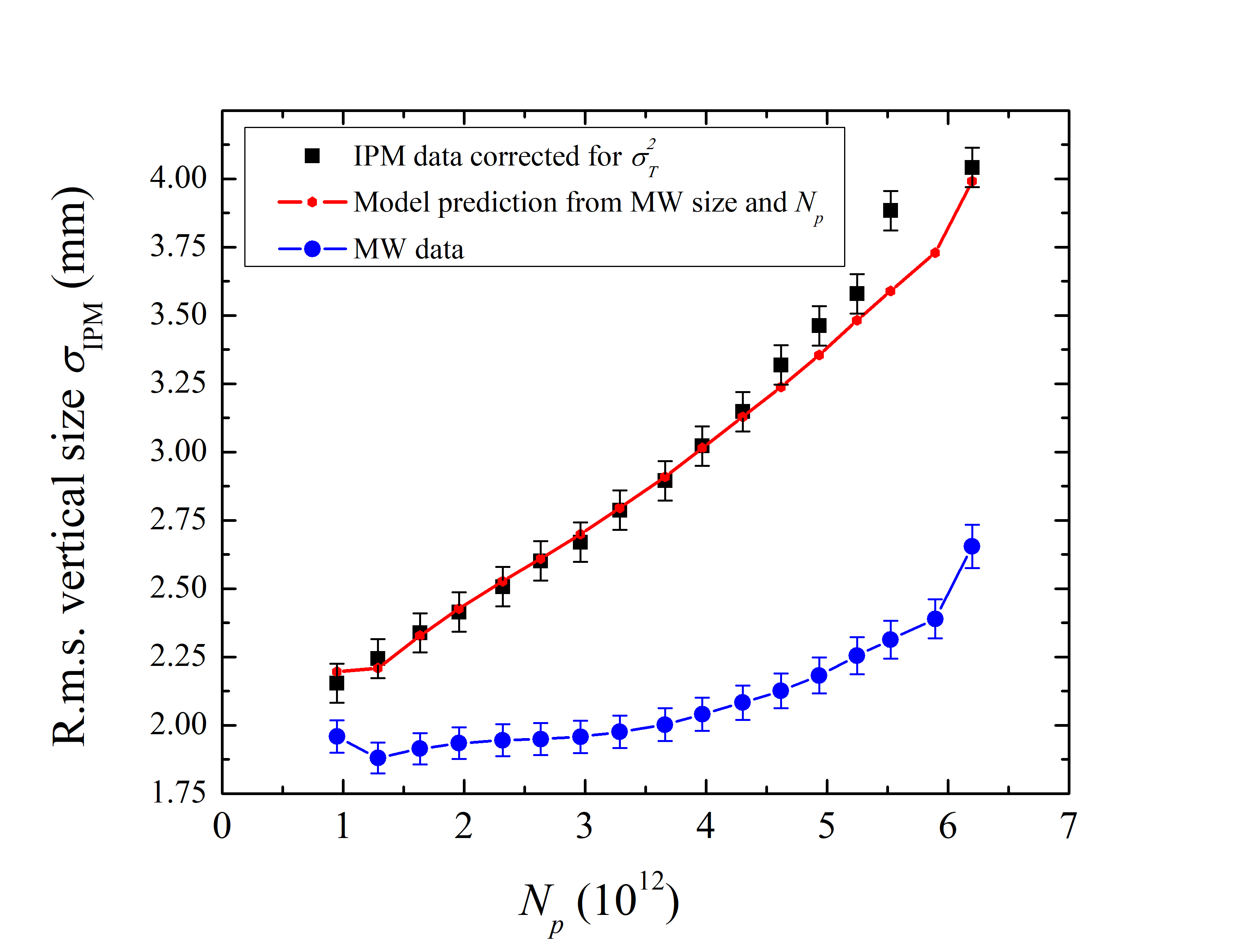}
\caption{The Fermilab Booster IPM data right before beam extraction $\sigma^*$ ($V_0=24$kV, $D=103$mm, black squares) \cite{Shiltsev2020} vs the total proton beam intensity $N$. The theoretical predication of this paper's Eq.(\ref{equationX6}) (red line) is calculated using the initial beam sizes $\sigma_0$ as measured by the Multi-Wires emittance monitor (blue line). The measured IPM rms sizes $\sigma_m$ are corrected for the intensity independent smearing $\sigma^*=\sqrt{\sigma_m^2(t) - \sigma_T^2}$, with $\sigma_T^2=2.7$ mm$^2$.} 
\label{Fig.3}
\end{figure}

The Booster proton beam dynamics is quite complicated; the beam emittance growth and particle losses during the acceleration \cite{Shiltsev2020a} are of serious operational concerns as they limit high power operation of the entire Fermilab complex of accelerators for high energy neutrino physics \cite{ShiltsevMPLA, Convery}. Fast diagnostics of the Booster proton beam size and emittances is of critical importance. 

There are two types of instruments to measure beam sizes and therefore, transverse emittances, in the Booster $-$ the Multi-Wires (MWs) and two ionization profile monitors. The Multi-Wires are  intercepting devices installed in the Booster extraction beamline. When the proton beam strikes individual wire (there are 48 wires spaced by 1 mm in each MW), secondary electrons create a current in the wires which is amplified to produce the profile. The MWs beam size measurement data are presumed to be intensity independent and accurate to some 2-3\%. Two IPMs $-$ vertical and horizontal $-$ operate in the ion collection mode and  report the average rms beam sizes (determined by the Gaussian fits of the profiles) every turn. As discussed above, the proton beam space charge fields lead to expansion of the transverse distribution in the IPMs and their outcomes are dependent on the proton beam intensity, as shown in \ref{Fig.3}. Therefore, a correction is needed to determine the actual rms proton beam sizes. 

For a majority of practical IPM applications, the most important outcome is the knowledge of the rms sizes of high energy beams with 5-10\% accuracy on a turn-by-turn or comparable time scale. That would correspond to about 10-20\% error in the beam emittances - a level comparable with capabilities of other, usually much slower types of beam size diagnostics which then can be used for cross--calibration \cite{moore2009beam, shiltsev2012accelerator, Roncarolo2005, Shiltsev2020}. 

Besides the effect of the initial velocities Eq.(\ref{Sigma5}), the IPM intensity independent profile smearing can be caused by: a) finite separation $\Delta$ between the individual IPM charge collection strips $\sigma_T \simeq \Delta/\sqrt{12}$; b) angular misalignment $\theta$ of the long and narrow strips with respect to the high energy beam trajectory $\sigma_T \simeq \theta L$, where $L$ is the strip length; c) charging of dielectric material in between the strips \cite{lebedev2020} or strip-to-stripe capacitive cross talk; d) non-uniformity of the extraction electric field in the operational IPM aperture  $\sigma_T \simeq \sigma_0 | dE_x/dx | / E_y)=\sigma_0 | dE_y/dy | / E_y)$. The latter effect is usually minimized by proper electro-mechanical design.

All the above effects are monitor-specific and the easiest way to account for them is cross-calibration of low intensity beam sizes measured by the IPM $\sigma_m$ and by another instrument $\sigma_{MW}$. In that case, the desired rms instrumental smearing can be found as: 
\begin{equation}
\sigma_T^2 = \lim_{N \rightarrow \, 0} \Big( \sigma_m^2(N) - \sigma_{MW}^{2}(N) \Big) \, .
\label{Sigma6}
\end{equation} 
Comparison of the IPM data with the Booster MWs data at various beam intensities yields the intercept in Eq.(\ref{Sigma6}) of $\sigma_T^2=2.8\pm 0.1$ mm$^2$ \cite{Shiltsev2020}. 

At high intensity, the space-charge potential of the Booster proton beam is $U_{SC}\approx 18.2 \cdot (N/6 \cdot 10^{12})$[V]. Typical rms bunch length and bunch-to-bunch spacing in the Booster are $\tau_b\approx 2-3$ ns, $t_b=25-19$ ns. The characteristic times for the IPM with $D=103$, $V_0=24$ kV, considered in the previous Section, are $\tau_0 \approx 22$ ns,  $\tau_1 \approx 67$ ns (for $N= 6 \cdot 10^{12}$) and $\tau_2 \approx 110$ ns. Therefore, the beam profile expansion factor $h$ can be calculated by using Eq.(\ref{equationX6}) in which the original $\sigma_0$ is taken from the MW data and the beam-to-MCP distance is taken as $d\approx D/2=53$ mm. The numerical coefficient $4\sqrt{2}/3$ in Eq.(\ref{equationX6}) should be multiplied by $(1+t_b/\tau_0)$ to account for the bunched beam time structure. The resulting rms vertical IPM beam sizes are found to be in excellent agreement with the actual  IPM sizes $\sigma^*=\sqrt{\sigma_m^2(t) - \sigma_T^2}$ measured over a broad range of beam intensities as shown in Fig.(\ref{Fig.3}). 

Knowing $\sigma_T, N$ and the IPM extracting field $V_0/D$ one can reverse Eq.(\ref{equationX6}) and find the original proton beam $\sigma_0$ from the measured and corrected $\sigma^*$, e.g., following Eq.(\ref{R1}) described in Appendix \ref{A4}. Fig.\ref{Fig.4} illustrates the result of such analysis for the measured profiles of the Booster beam with $N=4.62 \cdot 10^{12}$ \cite{Shiltsev2020}. There, the red curve is for the rms vertical beam size $\sigma_m (t)$ as reported by the IPM; the line in violet represents the beam size after correction for the intensity independent smearing $\sqrt{\sigma_m^2(t) - \sigma_T^2}$; and, finally, the true proton rms beam size $\sigma_0$ was reconstructed following Eq. (\ref{R3}) and is represented by the green line. One can see that the overall beam size correction is about 15\% early in the Booster acceleration cycle when the rms beam size is about 6 mm. At the end of the cycle, with proton energy increased from 400 MeV to 8 GeV, the correction is almost by a factor of two and accounting for the space-charge expansion is the most important.   

\begin{figure}[htbp]
\centering
\includegraphics[width=1.0\linewidth]{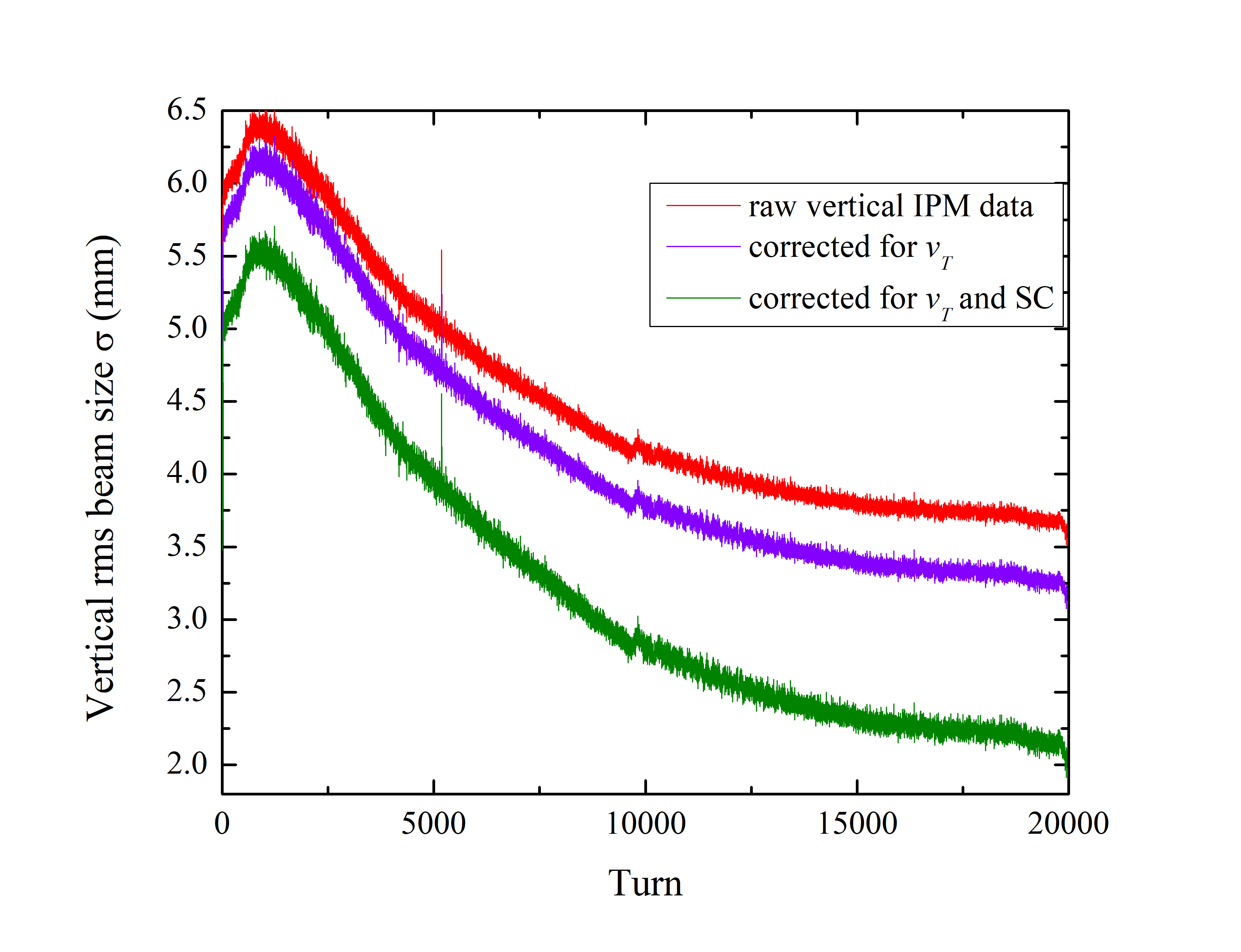}
\caption{An example of reconstruction of vertical rms proton beam size in the 33 ms (20000 turns) acceleration cycle of the Fermilab 8 GeV Booster synchtrotron with the total beam intensity of $N=4.6 \cdot 10^{12}$: time dependence of the original IPM data (red), the data corrected for smearing effects (violet) and the same data after additional correction for the space-charge expansion (green). } 
\label{Fig.4}
\end{figure}
\section{Conclusions}
\label{Discusission}
Ionization profile monitors are widely used in various types of particle accelerators for non-intercepting and fast beam profile measurements. As discussed in this paper, the major profile distortions in IPMs are due to expansion of slow ionization secondaries under the impact of the space-charge forces of the charged particle beam itself. The distortion is independent of the type of collected secondaries (different ions, electrons) and grows with an increase of beam intensity $N$ and IPM gap $D$ and a decrease of the beam size $\sigma_0$ and the IPM extracting voltage $V_0$ - see Eqs.(\ref{equationX6}) and (\ref{EQ6}).  However, the smearing effect due to significant initial kinetic energy $\mathcal E_i$ precludes operation of IPMs in the electron collection mode unless a strong external magnetic field is applied along with the IPM electric field. 

We have developed a model and an algorithm to account for the space-charge expansion in the ion collecting IPMs without an external magnetic field. Our theory goes significantly beyond previous analyses as it develops for the first time the proper functional dependence of the expansion on just few key IPM and beam parameters. The model verification against numerical particle tracking simulations results shows its excellent applicability. Contrary to previously considered phenomenological approximations which had several, up to four, numerical coefficients for the IPM profile space-charge induced expansion effects, the presented theory is concise and complete while precise. 

The rms beam size reconstruction  according to Eqs.(\ref{equationX6}) and (\ref{EQ6}) allows 5-10\% or better accuracy in determination of $\sigma_0$ from measured $\sigma_m$ and known beam intensity and IPM parameters $D$, $V_0$ and $d$ (distance from the beam orbit to the IPM MCP plate). 

Other, intensity independent instrumental errors $\sigma_T$ can easily be accounted for in quadrature if the IPM measurements are calibrated against other beam size diagnostics instruments \cite{strehl2006beam, Zimmermann, Bravin} at low beam intensities. The proposed algorithm, though simple and straightforward and addressing the most common operational needs, can not substitute for more sophisticated modeling and analysis if detailed knowledge of the high energy beam distribution (shape, tails, etc) is required. 

\section*{Acknowledgements} 
I would like to thank Jeff Eldred, Valery Kapin, Valery Lebedev, and Kiyomi Seiya for useful discussions, fruitful cooperation and valuable input. I am very grateful to Randy Thurman-Keup for detailed description of the operational Fermilab Booster IPMs, for providing Fig.\ref{Fig.1} and to Valerie Higgins and Karin Kemp for carefully reading through the manuscript and giving helpful feedback. 
Fermi National Accelerator Laboratory is operated by the Fermi Research Alliance, LLC under Contract No. DE-AC02-07CH11359 with the United States Department of Energy.
\\

\appendix
\section{Expansion of Gaussian beam profile}
\label{A1}
The electric force of the primary Gaussian beam is 
\begin{equation}
E^{\rm SC}_{(x,y)} = \frac{2J}{(4\pi \epsilon_0) v_p} \frac{(x,y)}{r^2} \Big( 1- {\rm exp} (-{r^2 \over {2 \sigma_0^2}}) \Big) \, , 
\label{EQ2}
\end{equation} 
where $J$ and $\sigma_0$ are the main (proton) beam current and rms transverse size, respectively, $v_p$ is its velocity, and $r^2=x(t)^2+y(t)^2$. IPMs usually operate with electric fields $E_y \sim O$(100-1000 V/mm) which significantly exceed the space-charge forces $E^{\rm SC} \sim O$(1-10 V/mm) and that makes the equation of motion in the $y$-plane trivial: 
\begin{equation}
    y(t) \approx \frac{ZeE_y}{2M} t^2.
\label{EQ3}
\end{equation}
Taking into account that the major part of the trajectory of the secondary particle lays outside the beam and, therefore, $r(t) \approx y(t)$, and that the initial coordinates are small compared to the average distance $d$ from the beam center to the IPM detector plane $(x_0, y_0) \ll d \approx D/2$ and one gets by substituting Eqs.(\ref{EQ2}, \ref{EQ3}) into Eq.(\ref{E1}): 
\begin{equation}
    x''(t) = \frac{x}{\tau_1^2} \frac{2\sigma_0^2}{y^2(t)} \Big( 1- {\rm exp} (-{r^2(t) \over {2 \sigma_0^2}}) \Big).
\label{EQ4}
\end{equation} 
where $\tau_1=\sqrt{M\sigma_0^2 / ZeU_{SC}}$ is a characteristic expansion time due to the space-charge, $U_{SC}=J/(4\pi \epsilon_0)v_p$. 
The solution of this second-order ordinary differential equation corresponding to initial conditions $x(0)=x_0$ and $x'(0)=0$ can be found via sequential approximations $x_{[0]}, x_{[1]}, x_{[2]}, ...x_{[n]}...$ each differing from the previous one by a term proportional to $(\tau_0/\tau_1)^{2n}$, where $\tau_0 = (2^{3/2} M\sigma_0/(ZeE_y)^{1/2}$ is a characteristic time needed for the secondary particle to get out of the beam under the force of external field $E_{\rm ext}$. Assuming that term to be small $\tau_0/\tau_1 \leq 1$, the first approximation yields: 
\begin{eqnarray}
x_{[1]}(t)=x_0 \cdot \Big[ 1+\frac{\tau_0^2}{\tau_1^2}\Big( \frac{t}{3\tau_0}\big(\Gamma ({1 \over 4}) - \Gamma ({1 \over 4}, {t^4 \over \tau_0^4})\big) - &&
\nonumber \\
- {1 \over 2}\sqrt{\pi} {\rm{erf}} ({t^2 \over \tau_0^2}) + \frac{\tau_0^2}{6t^2}(1- {\rm exp} (-{t^4 \over \tau_0^4})\Big) \Big] &. & 
\label{EQ5} 
\end{eqnarray}
This equation is linear with respect to $x_0$, therefore, the space-charge expansion in IPM results in {\it proportional magnification of the profile} of the distribution of the secondary particles. At the time when secondary particles reach the IPM detector $t=\tau_2=\sqrt{2dM/ZeE_y} \gg \tau_0$, such magnification results in the profile rms size of: 
\begin{equation}
    \sigma_m \approx \sigma_0  \cdot \Big( 1 + \frac{2\Gamma({1 \over 4})}{3} 
    \frac{U_{SC}}{E_y \sigma_0}  \sqrt{d \over \sigma_0} 
\Big) \, .
\label{EQ6}
\end{equation} 
This expression is very close to Eq.(\ref{equationX6}) as the Gamma-function $\Gamma({1 \over 4}) \approx 3.625$.  
\begin{figure}[htbp]
\centering
\includegraphics[width=0.9\linewidth]{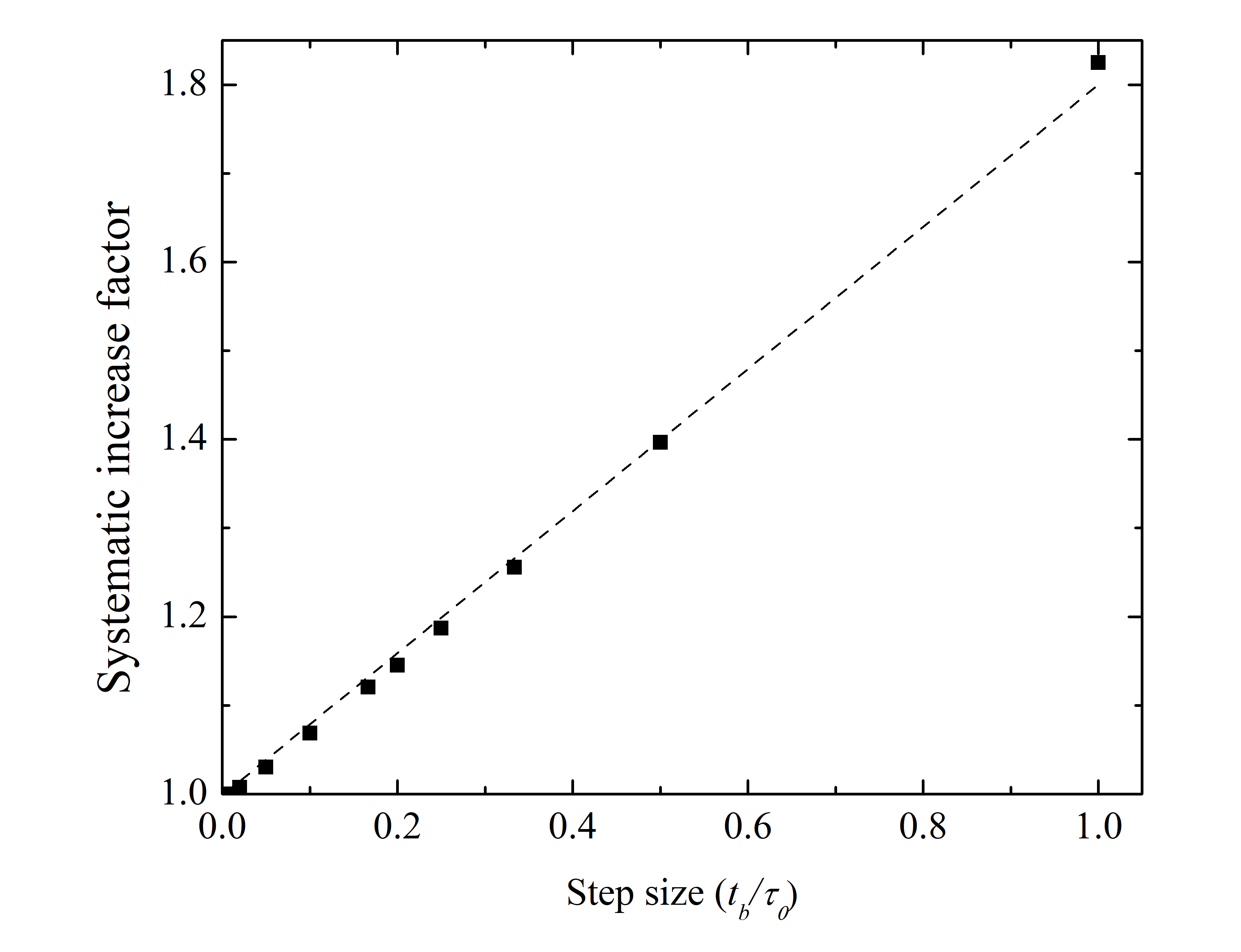}
\caption{Simulations of the space-charge expansion Eqs.(\ref{equationX1}, \ref{equationX2}). } 
\label{FigB}
\end{figure}

\section{Beams of short sparse bunches}
\label{A2}
In the case when proton bunches are short and spacing between them 
$t_b$ is comparable to $\tau_0$ and $\tau_2$, the main equations (\ref{equationX1}) and (\ref{equationX2}) can be substituted by corresponding sums of kicks due to individual bunches and solved numerically. As follows from the theory on numerical methods, such {\it Riemann sums} lead to the systematic increase of the integrals proportional to the
integration step size and the derivative of the function \cite{Sums}. 
Fig. \ref{FigB} shows results of summing functions Eqs.(\ref{equationX1},
\ref{equationX2}) at $t_n=n \times t_b \leq \tau_2=5\tau_0$ with 
{\it Wolfram Alpha} tool \cite{Wolfram} for different bunch spacing $t_b$. The sums are normalized to the value of continuous integral, which corresponds to $t_b \rightarrow 0$, and the dashed line represents the systematic increase factor of $(1+0.8\, t_b/\tau_0)$. The accuracy of this correction due to the time discretization is $\sim(t_b/ \tau_2)$, e.g. about 4\% at $t_b = 0.2 \tau_0$.  

\section{Averaging over Gaussian distributions}
\label{A3}

We start with modified Eq.(\ref{short2}) for the final position of the particle $x_f$ vs its initial position $x$:
\begin{equation}
x_f = x \Big[ 1+ {\lambda \over u^2} \Big( 1- {\rm exp} (-{u^2/2}) \Big) \Big] \, ,  
\label{B1}
\end{equation} 
where $\lambda=(4 \kappa N_p)/(\sigma_0^2)$ and $u=r/\sigma_0$. To get the mean square of the final position distribution, one needs to calculate the integral over the initial Gaussian distribution $\rho(x,y)$: 
\begin{equation}
\sigma_f^2=\iint \rho(x,y)dxdy \, x^2 \Big[ 1+ {\lambda \over u^2} \Big( 1- e^{-u^2/2} \Big) \Big]^2 \, .  
\label{B2}
\end{equation} 

Given similar distributions in $x$ and $y$, the integrals in both planes are the same, one can sum them up and switch to polar coordinates $\rho(x,y)dxdy=2\pi\rho(u)udu$ to end up with: 
\begin{equation}
\sigma_f^2={\sigma_0^2 \over 2} \int_{0}^{\infty} e^{-u^2/2}du \, u^3 \Big[ 1+ {\lambda \over u^2} \Big( 1- e^{-u^2/2} \Big) \Big]^2 \, .  
\label{B3}
\end{equation}
The squared expression in the brackets can be opened up as $1+2f+f^2$, where $f=\lambda ( 1- e^{-u^2/2}) / u^2$, and  integrating each term separately yields the total of:
\begin{equation}
\sigma_f^2={\sigma_0^2 \over 2} \Big[ 2 + \lambda + \lambda^2 {\rm ln}(2 / \sqrt{3}) \Big] \, . 
\label{B4}
\end{equation}

Now, we note that $\lambda$ scales with number of primary particles (protons) which impacted the secondary particle (ion). As the secondaries are created by the primaries, the former are affected only by the protons passing by {\it after} the act of ionization. For the Gaussian distribution of the proton current in time $t$, this time-dependent impact is proportional to $N_p (1+{\rm erfc}(t/\sigma_t  \sqrt{2}))/2$, where $\sigma_t$ is the rms pulse duration. Corresponding time-averaging of the three terms in Eq.(B4) over the proton pulse with weight ${\rm exp} (-t^2/2 \sigma_t^2)$ yields
\begin{equation}
\sigma_f^2=\sigma_0^2 \Big[ 1 + {1 \over 4} \lambda + {1 \over 6} \lambda^2 {\rm ln}(2 / \sqrt{3}) \Big] \, , 
\label{B5}
\end{equation}
that is Eq.(\ref{short3}). 

\section{Finding Original RMS Sizes}
\label{A4}

There are many ways to find $\sigma_0$ from $\sigma_m$, i.e., to solve equations of the type of (\ref{equationX6}) and (\ref{EQ6}): 
\begin{equation}
    \sigma_m = \sigma_0 \cdot \Big(
1+ {c \over \sigma_0^{3/2}} \Big) \, ,
\label{R1}
\end{equation} 
where in our case $c=(8\sqrt[4]{2}/ 3) (U_{SC} D \sqrt{d}/V_0)$. There is an exact solution of the cubic equation (\ref{R1}) but it is quite complicated and lengthy mathematical expression. Slightly easier and straightforward is an iterative approximation $\sigma_0^{[n]}$: 
\begin{equation}
    \sigma_0^{[n]} = \frac{\sigma_0^{[n-1]}} {1+ {c/(\sigma_0^{[n-1]})^{3/2}}} \, ,
\label{R2}
\end{equation}
that starts at $\sigma_0^{[1]}=\sigma_m$. Three to four iterations are usually enough to get satisfactorily accuracy. 

Analysing the IPM data with modest expansion $h=\sigma_m/\sigma_0 < 1+1/\sqrt[3]{2}$ we found even simpler practical algorithm that is good within 5\% : 
\begin{equation}
    \sigma_0 \approx \frac {\sigma_m}{1+ { (c+c^2/2) / \sigma_m^{3/2}}} \, .
\label{R3}
\end{equation}


\begin{thebibliography}{22}%
\makeatletter
\providecommand \@ifxundefined [1]{%
 \@ifx{#1\undefined}
}%
\providecommand \@ifnum [1]{%
 \ifnum #1\expandafter \@firstoftwo
 \else \expandafter \@secondoftwo
 \fi
}%
\providecommand \@ifx [1]{%
 \ifx #1\expandafter \@firstoftwo
 \else \expandafter \@secondoftwo
 \fi
}%
\providecommand \natexlab [1]{#1}%
\providecommand \enquote  [1]{``#1''}%
\providecommand \bibnamefont  [1]{#1}%
\providecommand \bibfnamefont [1]{#1}%
\providecommand \citenamefont [1]{#1}%
\providecommand \href@noop [0]{\@secondoftwo}%
\providecommand \href [0]{\begingroup \@sanitize@url \@href}%
\providecommand \@href[1]{\@@startlink{#1}\@@href}%
\providecommand \@@href[1]{\endgroup#1\@@endlink}%
\providecommand \@sanitize@url [0]{\catcode `\\12\catcode `\$12\catcode
  `\&12\catcode `\#12\catcode `\^12\catcode `\_12\catcode `\%12\relax}%
\providecommand \@@startlink[1]{}%
\providecommand \@@endlink[0]{}%
\providecommand \url  [0]{\begingroup\@sanitize@url \@url }%
\providecommand \@url [1]{\endgroup\@href {#1}{\urlprefix }}%
\providecommand \urlprefix  [0]{URL }%
\providecommand \Eprint [0]{\href }%
\providecommand \doibase [0]{http://dx.doi.org/}%
\providecommand \selectlanguage [0]{\@gobble}%
\providecommand \bibinfo  [0]{\@secondoftwo}%
\providecommand \bibfield  [0]{\@secondoftwo}%
\providecommand \translation [1]{[#1]}%
\providecommand \BibitemOpen [0]{}%
\providecommand \bibitemStop [0]{}%
\providecommand \bibitemNoStop [0]{.\EOS\space}%
\providecommand \EOS [0]{\spacefactor3000\relax}%
\providecommand \BibitemShut  [1]{\csname bibitem#1\endcsname}%
\let\auto@bib@innerbib\@empty

\bibitem [{\citenamefont {Hornstra}\ and\ \citenamefont
  {DeLuca}(1967)}]{hornstra1967nondestructive}%
  \BibitemOpen
  \bibfield  {author} {\bibinfo {author} {\bibfnamefont {F.}~\bibnamefont
  {Hornstra}}\ and\ \bibinfo {author} {\bibfnamefont {W.}~\bibnamefont
  {DeLuca}},\ }in\ \href@noop {} {\emph {\bibinfo {booktitle} {Proceedings VI
  Intl. Conf. on High Energy Accel. (Cambridge, MA)}}}\ (\bibinfo {year} {Sept.
  1967})\ pp.\ \bibinfo {pages} {374--377}\BibitemShut {NoStop}%
\bibitem [{\citenamefont {Weisberg}\ \emph {et~al.}(1983)\citenamefont
  {Weisberg}, \citenamefont {Gill}, \citenamefont {Ingrassia},\ and\
  \citenamefont {Rodger}}]{weisberg1983ionization}%
  \BibitemOpen
  \bibfield  {author} {\bibinfo {author} {\bibfnamefont {H.}~\bibnamefont
  {Weisberg}}, \bibinfo {author} {\bibfnamefont {E.}~\bibnamefont {Gill}},
  \bibinfo {author} {\bibfnamefont {P.}~\bibnamefont {Ingrassia}}, \ and\
  \bibinfo {author} {\bibfnamefont {E.}~\bibnamefont {Rodger}},\ }\href@noop {}
  {\bibfield  {journal} {\bibinfo  {journal} {IEEE Transactions on Nuclear
  Science}\ }\textbf {\bibinfo {volume} {30}},\ \bibinfo {pages} {2179}
  (\bibinfo {year} {1983})}\BibitemShut {NoStop}%
\bibitem [{\citenamefont {Hochadel}\ \emph {et~al.}(1994)\citenamefont
  {Hochadel}, \citenamefont {Albrecht}, \citenamefont {Grieser}, \citenamefont
  {Habs}, \citenamefont {Schwalm}, \citenamefont {Szmola},\ and\ \citenamefont
  {Wolf}}]{hochadel1994residual}%
  \BibitemOpen
  \bibfield  {author} {\bibinfo {author} {\bibfnamefont {B.}~\bibnamefont
  {Hochadel}}, \bibinfo {author} {\bibfnamefont {F.}~\bibnamefont {Albrecht}},
  \bibinfo {author} {\bibfnamefont {M.}~\bibnamefont {Grieser}}, \bibinfo
  {author} {\bibfnamefont {D.}~\bibnamefont {Habs}}, \bibinfo {author}
  {\bibfnamefont {D.}~\bibnamefont {Schwalm}}, \bibinfo {author} {\bibfnamefont
  {E.}~\bibnamefont {Szmola}}, \ and\ \bibinfo {author} {\bibfnamefont
  {A.}~\bibnamefont {Wolf}},\ }\href@noop {} {\bibfield  {journal} {\bibinfo
  {journal} {Nuclear Instruments and Methods in Physics Research Section A:
  Accelerators, Spectrometers, Detectors and Associated Equipment}\ }\textbf
  {\bibinfo {volume} {343}},\ \bibinfo {pages} {401} (\bibinfo {year}
  {1994})}\BibitemShut {NoStop}%
\bibitem [{\citenamefont {Anne}\ \emph {et~al.}(1993)\citenamefont {Anne},
  \citenamefont {Georget}, \citenamefont {Hue}, \citenamefont {Tribouillard},\
  and\ \citenamefont {Vignet}}]{anne1993noninterceptive}%
  \BibitemOpen
  \bibfield  {author} {\bibinfo {author} {\bibfnamefont {R.}~\bibnamefont
  {Anne}}, \bibinfo {author} {\bibfnamefont {Y.}~\bibnamefont {Georget}},
  \bibinfo {author} {\bibfnamefont {R.}~\bibnamefont {Hue}}, \bibinfo {author}
  {\bibfnamefont {C.}~\bibnamefont {Tribouillard}}, \ and\ \bibinfo {author}
  {\bibfnamefont {J.~L.}\ \bibnamefont {Vignet}},\ }\href@noop {} {\bibfield
  {journal} {\bibinfo  {journal} {Nuclear Instruments and Methods in Physics
  Research Section A: Accelerators, Spectrometers, Detectors and Associated
  Equipment}\ }\textbf {\bibinfo {volume} {329}},\ \bibinfo {pages} {21}
  (\bibinfo {year} {1993})}\BibitemShut {NoStop}%
\bibitem [{\citenamefont {Connolly}\ \emph {et~al.}(2000)\citenamefont
  {Connolly}, \citenamefont {Michnoff}, \citenamefont {Moore}, \citenamefont
  {Shea},\ and\ \citenamefont {Tepikian}}]{connolly2000beam}%
  \BibitemOpen
  \bibfield  {author} {\bibinfo {author} {\bibfnamefont {R.}~\bibnamefont
  {Connolly}}, \bibinfo {author} {\bibfnamefont {R.}~\bibnamefont {Michnoff}},
  \bibinfo {author} {\bibfnamefont {T.}~\bibnamefont {Moore}}, \bibinfo
  {author} {\bibfnamefont {T.}~\bibnamefont {Shea}}, \ and\ \bibinfo {author}
  {\bibfnamefont {S.}~\bibnamefont {Tepikian}},\ }\href@noop {} {\bibfield
  {journal} {\bibinfo  {journal} {Nuclear Instruments and Methods in Physics
  Research Section A: Accelerators, Spectrometers, Detectors and Associated
  Equipment}\ }\textbf {\bibinfo {volume} {443}},\ \bibinfo {pages} {215}
  (\bibinfo {year} {2000})}\BibitemShut {NoStop}%
\bibitem [{\citenamefont {Jansson}\ \emph {et~al.}(2006)\citenamefont
  {Jansson}, \citenamefont {Fitzpatrick}, \citenamefont {Bowie}, \citenamefont
  {Kwarciany}, \citenamefont {Lundberg}, \citenamefont {Slimmer}, \citenamefont
  {Valerio},\ and\ \citenamefont {Zagel}}]{jansson2006tevatron}%
  \BibitemOpen
  \bibfield  {author} {\bibinfo {author} {\bibfnamefont {A.}~\bibnamefont
  {Jansson}}, \bibinfo {author} {\bibfnamefont {T.}~\bibnamefont
  {Fitzpatrick}}, \bibinfo {author} {\bibfnamefont {K.}~\bibnamefont {Bowie}},
  \bibinfo {author} {\bibfnamefont {R.}~\bibnamefont {Kwarciany}}, \bibinfo
  {author} {\bibfnamefont {C.}~\bibnamefont {Lundberg}}, \bibinfo {author}
  {\bibfnamefont {D.}~\bibnamefont {Slimmer}}, \bibinfo {author} {\bibfnamefont
  {L.}~\bibnamefont {Valerio}}, \ and\ \bibinfo {author} {\bibfnamefont
  {J.}~\bibnamefont {Zagel}},\ }in\ \href@noop {} {\emph {\bibinfo {booktitle}
  {AIP Conference Proceedings}}},\ Vol.\ \bibinfo {volume} {868}\ (\bibinfo
  {organization} {American Institute of Physics},\ \bibinfo {year} {2006})\
  pp.\ \bibinfo {pages} {159--167}\BibitemShut {NoStop}%
\bibitem [{\citenamefont {Levasseur}\ \emph {et~al.}(2017)\citenamefont
  {Levasseur}, \citenamefont {Dehning}, \citenamefont {Gibson}, \citenamefont
  {Sandberg}, \citenamefont {Sapinski}, \citenamefont {Sato}, \citenamefont
  {Schneider},\ and\ \citenamefont {Storey}}]{levasseur2017development}%
  \BibitemOpen
  \bibfield  {author} {\bibinfo {author} {\bibfnamefont {S.}~\bibnamefont
  {Levasseur}}, \bibinfo {author} {\bibfnamefont {B.}~\bibnamefont {Dehning}},
  \bibinfo {author} {\bibfnamefont {S.}~\bibnamefont {Gibson}}, \bibinfo
  {author} {\bibfnamefont {H.}~\bibnamefont {Sandberg}}, \bibinfo {author}
  {\bibfnamefont {M.}~\bibnamefont {Sapinski}}, \bibinfo {author}
  {\bibfnamefont {K.}~\bibnamefont {Sato}}, \bibinfo {author} {\bibfnamefont
  {G.}~\bibnamefont {Schneider}}, \ and\ \bibinfo {author} {\bibfnamefont
  {J.}~\bibnamefont {Storey}},\ }\href@noop {} {\bibfield  {journal} {\bibinfo
  {journal} {Journal of instrumentation}\ }\textbf {\bibinfo {volume} {12}},\
  \bibinfo {pages} {C02050} (\bibinfo {year} {2017})}\BibitemShut {NoStop}%
\bibitem [{\citenamefont {Moore}\ \emph {et~al.}(2009)\citenamefont {Moore},
  \citenamefont {Jansson},\ and\ \citenamefont {Shiltsev}}]{moore2009beam}%
  \BibitemOpen
  \bibfield  {author} {\bibinfo {author} {\bibfnamefont {R.~S.}\ \bibnamefont
  {Moore}}, \bibinfo {author} {\bibfnamefont {A.}~\bibnamefont {Jansson}}, \
  and\ \bibinfo {author} {\bibfnamefont {V.}~\bibnamefont {Shiltsev}},\
  }\href@noop {} {\bibfield  {journal} {\bibinfo  {journal} {Journal of
  Instrumentation}\ }\textbf {\bibinfo {volume} {4}},\ \bibinfo {pages}
  {P12018} (\bibinfo {year} {2009})}\BibitemShut {NoStop}%
\bibitem [{\citenamefont {Benedetti}\ \emph {et~al.}(2020)\citenamefont
  {Benedetti}, \citenamefont {Abbon}, \citenamefont {Belloni}, \citenamefont
  {Coulloux}, \citenamefont {Gougnaud}, \citenamefont {Lahonde-Hamdoun},
  \citenamefont {Le~Bourlout}, \citenamefont {Mariette}, \citenamefont
  {Marroncle}, \citenamefont {Mols} \emph {et~al.}}]{benedetti2020design}%
  \BibitemOpen
  \bibfield  {author} {\bibinfo {author} {\bibfnamefont {F.}~\bibnamefont
  {Benedetti}}, \bibinfo {author} {\bibfnamefont {P.}~\bibnamefont {Abbon}},
  \bibinfo {author} {\bibfnamefont {F.}~\bibnamefont {Belloni}}, \bibinfo
  {author} {\bibfnamefont {G.}~\bibnamefont {Coulloux}}, \bibinfo {author}
  {\bibfnamefont {F.}~\bibnamefont {Gougnaud}}, \bibinfo {author}
  {\bibfnamefont {C.}~\bibnamefont {Lahonde-Hamdoun}}, \bibinfo {author}
  {\bibfnamefont {P.}~\bibnamefont {Le~Bourlout}}, \bibinfo {author}
  {\bibfnamefont {Y.}~\bibnamefont {Mariette}}, \bibinfo {author}
  {\bibfnamefont {J.}~\bibnamefont {Marroncle}}, \bibinfo {author}
  {\bibfnamefont {J.}~\bibnamefont {Mols}},  \emph {et~al.},\ }in\ \href@noop
  {} {\emph {\bibinfo {booktitle} {EPJ Web of Conferences}}},\ Vol.\ \bibinfo
  {volume} {225}\ (\bibinfo {organization} {EDP Sciences},\ \bibinfo {year}
  {2020})\ p.\ \bibinfo {pages} {01009}\BibitemShut {NoStop}%
\bibitem [{\citenamefont {Strehl}(2006)}]{strehl2006beam}%
  \BibitemOpen
  \bibfield  {author} {\bibinfo {author} {\bibfnamefont {P.}~\bibnamefont
  {Strehl}},\ }\href@noop {} {\emph {\bibinfo {title} {Beam instrumentation and
  diagnostics}}},\ Vol.\ \bibinfo {volume} {120}\ (\bibinfo  {publisher}
  {Springer},\ \bibinfo {year} {2006})\BibitemShut {NoStop}%
\bibitem [{\citenamefont {Wittenburg}(2013)}]{wittenburg2013specific}%
  \BibitemOpen
  \bibfield  {author} {\bibinfo {author} {\bibfnamefont {K.}~\bibnamefont
  {Wittenburg}},\ }\href@noop {} {\bibfield  {journal} {\bibinfo  {journal}
  {arXiv preprint arXiv:1303.6767}\ } (\bibinfo {year} {2013})}\BibitemShut
  {NoStop}%
\bibitem [{\citenamefont {Vilsmeier}\ \emph {et~al.}(2019)\citenamefont
  {Vilsmeier}, \citenamefont {Sapinski},\ and\ \citenamefont
  {Singh}}]{vilsmeier2019space}%
  \BibitemOpen
  \bibfield  {author} {\bibinfo {author} {\bibfnamefont {D.}~\bibnamefont
  {Vilsmeier}}, \bibinfo {author} {\bibfnamefont {M.}~\bibnamefont {Sapinski}},
  \ and\ \bibinfo {author} {\bibfnamefont {R.}~\bibnamefont {Singh}},\
  }\href@noop {} {\bibfield  {journal} {\bibinfo  {journal} {Physical Review
  Accelerators and Beams}\ }\textbf {\bibinfo {volume} {22}},\ \bibinfo {pages}
  {052801} (\bibinfo {year} {2019})}\BibitemShut {NoStop}%
\bibitem [{\citenamefont {{Ionisation Profile Monitor Simulation Kickoff
  Workshop (CERN, March 3-4, 2016)}}()}]{workshop2016}%
  \BibitemOpen
  \bibfield  {author} {\bibinfo {author} {\bibnamefont {{Ionisation Profile
  Monitor Simulation Kickoff Workshop (CERN, March 3-4, 2016)}}},\ }\href@noop
  {} {}\bibinfo {howpublished} {\url{https://indico.cern.ch/event/491615/}},\
  \bibinfo {note} {accessed: March 3, 2020}\BibitemShut {NoStop}%
\bibitem [{\citenamefont {{Workshop on Simulations, Design and Operational
  Experience of Ionisation Profile Monitors (GSI, May 21-24,
  2017)}}()}]{workshop2017}%
  \BibitemOpen
  \bibfield  {author} {\bibinfo {author} {\bibnamefont {{Workshop on
  Simulations, Design and Operational Experience of Ionisation Profile Monitors
  (GSI, May 21-24, 2017)}}},\ }\href@noop {} {}\bibinfo {howpublished}
  {\url{https://indico.gsi.de/event/5366/}},\ \bibinfo {note} {accessed: March
  3, 2020}\BibitemShut {NoStop}%
\bibitem [{\citenamefont {Thern}(1987)}]{thern1987space}%
  \BibitemOpen
  \bibfield  {author} {\bibinfo {author} {\bibfnamefont {R.}~\bibnamefont
  {Thern}},\ }in\ \href@noop {} {\emph {\bibinfo {booktitle} {1987 IEEE
  Particle Accelerator Conf.(PAC'87), Washington, DC, 16-19 May 1987}}}\
  (\bibinfo {organization} {IEEE},\ \bibinfo {year} {1987})\ pp.\ \bibinfo
  {pages} {646--648}\BibitemShut {NoStop}%
\bibitem{Graves} W. Graves, 
Nuclear Instruments and Methods in Physics Research Section A: Accelerators, Spectrometers, Detectors and Associated Equipment {\bf 364}, no.1, pp.19-26 (1995).
\bibitem [{\citenamefont {Amundson}\ \emph {et~al.}(2003)\citenamefont
  {Amundson}, \citenamefont {Lackey}, \citenamefont {Spentzouris},
  \citenamefont {Jungman},\ and\ \citenamefont
  {Spentzouris}}]{amundson2003calibration}%
  \BibitemOpen
  \bibfield  {author} {\bibinfo {author} {\bibfnamefont {J.}~\bibnamefont
  {Amundson}}, \bibinfo {author} {\bibfnamefont {J.}~\bibnamefont {Lackey}},
  \bibinfo {author} {\bibfnamefont {P.}~\bibnamefont {Spentzouris}}, \bibinfo
  {author} {\bibfnamefont {G.}~\bibnamefont {Jungman}}, \ and\ \bibinfo
  {author} {\bibfnamefont {L.}~\bibnamefont {Spentzouris}},\ }\href@noop {}
  {\bibfield  {journal} {\bibinfo  {journal} {Physical Review Special
  Topics-Accelerators and Beams}\ }\textbf {\bibinfo {volume} {6}},\ \bibinfo
  {pages} {102801} (\bibinfo {year} {2003})}\BibitemShut {NoStop}%
\bibitem{Reiser}  M.Reiser, {\it Theory and design of charged particle beams} (John Wiley \& Sons, 2008).
\bibitem{Morse}  P.Morse, H.Feshbach, {\it Methods of Theoretical Physics} (Mc Graw-Hill Book Co, 1953).
\bibitem [{\citenamefont {Bakker}\ and\ \citenamefont
  {Segre}(1951)}]{bakker1951stopping}%
  \BibitemOpen
  \bibfield  {author} {\bibinfo {author} {\bibfnamefont {C.}~\bibnamefont
  {Bakker}}\ and\ \bibinfo {author} {\bibfnamefont {E.}~\bibnamefont {Segre}},\
  }\href@noop {} {\bibfield  {journal} {\bibinfo  {journal} {Physical Review}\
  }\textbf {\bibinfo {volume} {81}},\ \bibinfo {pages} {489} (\bibinfo {year}
  {1951})}\BibitemShut {NoStop}%
\bibitem{Dimopoulou2005} C. Dimopoulou, et al., Journal of Physics B: Atomic, Molecular and Optical Physics {\bf 38 (5)}, 593 (2005). 
\bibitem{Egberts2012} J. Egberts, {\it IFMIF-LIPAc Beam Diagnostics. Profiling and Loss Monitoring Systems.} PhD Diss., Universite Paris Sud (2012).
\bibitem {Shiltsev2020} V.Shiltsev, J.Eldred, V. Lebedev , K.Seiya, {\it Studies of Beam Intensity Effects in Fermilab Booster Synchrotron. Part II: Beam Emittance Evolution}, Preprint FERMILAB-TM-2741 (2020). 
\bibitem{Booster} E.L.Hubbard (ed.), {\it Booster Synchrotron}, Preprint FERMILAB-TM-405 (1973);the Booster injection energy was increased to 400 MeV in 1990, see {it Fermilab Linac Upgrade Conceptual Design,} Fermilab Memo, Fermilab LU Conceptual Design (1989).
\bibitem{BoosterBook} {\it Booster Rookie Book}, {\url{https://operations.fnal.gov/rookie_books/Booster_V4.1.pdf}}
\bibitem {Shiltsev2020a} V.Shiltsev, J.Eldred, V. Lebedev , K.Seiya, {\it Studies of Beam Intensity Effects in Fermilab Booster Synchrotron. Part I: Introduction; Tune and Chromaticity Scans of Beam Losses}, Preprint FERMILAB-TM-2740 (2020). 
\bibitem{ShiltsevMPLA} V.Shiltsev, Modern Physics Letters A, {\bf 32(16)}, 1730012 (2017)
\bibitem{Convery} M. Convery, M.Lindgren, S.Nagaitsev, V.Shiltsev, Fermilab Accelerator Complex:
Status and Improvement Plans, Preprint FERMILAB-TM-2693 (2018).
\bibitem {shiltsev2012accelerator} V.Lebedev, V.Shiltsev (eds.), {\it Accelerator physics at the Tevatron collider} (Springer, 2012)
\bibitem{Roncarolo2005} F. Roncarolo,{\it Accuracy of the transverse emittance measurements of the CERN Large Hadron Collider},  CERN-THESIS-2005-082, PhD Diss., Milan Polytechnic (2005).  
\bibitem{lebedev2020} V.Lebedev, private  communication (2020). 
\bibitem{Zimmermann}  M. Minty, F. Zimmermann,  {\it Measurement and control of charged particle beams}, (Springer, 2004). 
\bibitem [{\citenamefont {Bravin}(2020)}]{Bravin}%
  \BibitemOpen
  \bibfield  {author} {\bibinfo {author} {\bibfnamefont {E.}~\bibnamefont
  {Bravin}},\ }\href@noop {} {\bibfield  {journal} {\bibinfo  {journal}
  {arXiv preprint arXiv:2005.07400}\ } (\bibinfo {year} {2020})}\BibitemShut
  {NoStop}%
\bibitem{Sums} D.Hughes-Hallett, W. McCullum, et al., {\it Calculus (4th ed.)} (Wiley, 2005). 
\bibitem{Wolfram} {\it Wolfram Alpha}, https://www.wolframalpha.com (accessed: April 29, 2020).%
%
\end{thebibliography}
%

\end{document}